# Design of a Double Spoke Cavity for C-ADS

JIANG Tian-Cai (蒋天才)[1,2] Huang Yu-Lu (黄玉璐)[1,2] HE Yuan(何源)[1] LU Xiang-Yang (鲁向阳) [1,3] Zhang Sheng-Hu (张生虎)[1]

[1] Institute of Modern Physics, Chinese Academy of Sciences, Lanzhou 730000, China

[2] University of Chinese Academy of Science, Beijing 100049, China

[3] State Key Laboratory of Nuclear Physics and Technology, Peking University, Beijing 100871, China

**Abstract:** The design of a 325MHz $\beta$=0.37 superconducting double spoke cavity for China ADS has been finished at Institute of Modern Physics. In this paper, the optimization of the spoke cavity is described in detail. The main goal in the EM design is to decrease the normalized surface field. We use the elliptic column instead of cylinder to decrease the normalized magnetic field. The electromagnetic simulation gives the optimum parameters *Ep/Eacc* of 4.39 and *Bp/Eacc* of 7.15mT/(MV/m), which meet the requirement of C-ADS. The mechanical properties were also studied.

**Key words:** double spoke, superconducting cavity, C-ADS, medium β, linac

**PACS:** 29.20.Ej

## 1 Introduction

The superconducting spoke cavity is based on one or more loaded structures where each loading element supports a TEM mode. So far several spoke cavities have been developed and tested for low and medium $\beta$ particle [1], and the spoke cavity is also studying for high velocity application [2]. While no spoke cavity is already in use in operating accelerators, they are under consideration for a number of them. Use of spoke cavities is planning in the main driver accelerator of the Project X [3] and the European spallation source [4]. Applications for the ADS (Accelerator Driven System) for nuclear waste transmutation or for tritium production also propose the use of spoke cavities [5].

Based on the scheme of China ADS [6], we have done researches on spoke cavities and designed an optimal $\beta_0$=0.37 double spoke cavity as an alternative option for spoke040.

## 2 EM design

The main goal in the EM design of superconducting cavity is to get a higher accelerating gradient and a lower heat load, which are determined by a lower peak surface fields (*Ep/Eacc* and *Bp/Eacc*) and a higher $G*R/Q_0$ ($G$ is the geometrical factor, $R$ is the shunt impedance and $Q_0$ is the quality factor). According to the state of the art, we require that the maximum peak surface electric and magnetic field are under 35MV/m and 70mT. For our application, the accelerating gradient is 8MV/m, hence the normalized electric field

* Supported by National Natural Science Foundation of China (91026004)

1) E-mail: jiangtiancai@impcas.ac.cn

*Ep/Eacc* and magnetic field *Bp/Eacc* is less than 4.38 and 8.75 respectively. In most cases, the performance limitation in a spoke cavity is the thermal-magnetic quench with little or no field emission [1], which leads us to take more effort to minimize *Bp/Eacc*.

This work has been done using the 3 dimensional electromagnetic design software CST Microwave Studio (MWS). Figure 1 shows a cut-away view of the double spoke cavity model in MWS and the main geometric parameters used for the optimization.

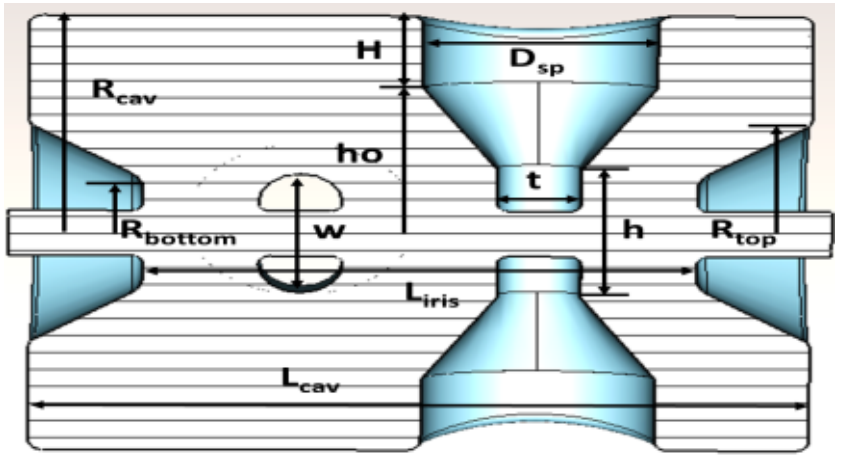

Fig.1. Cut-away view of double spoke cavity model in MWS and main parameters.

## 2.1 Dependence of RF properties on geometric parameters

**Spoke base and cavity length:** At first, we fix the cavity length (*Lcav*) and change the spoke base diameter (*Dsp*). The variation of the normalized field and shunt impedance with *Dsp /Lcav* is shown in figure 2.

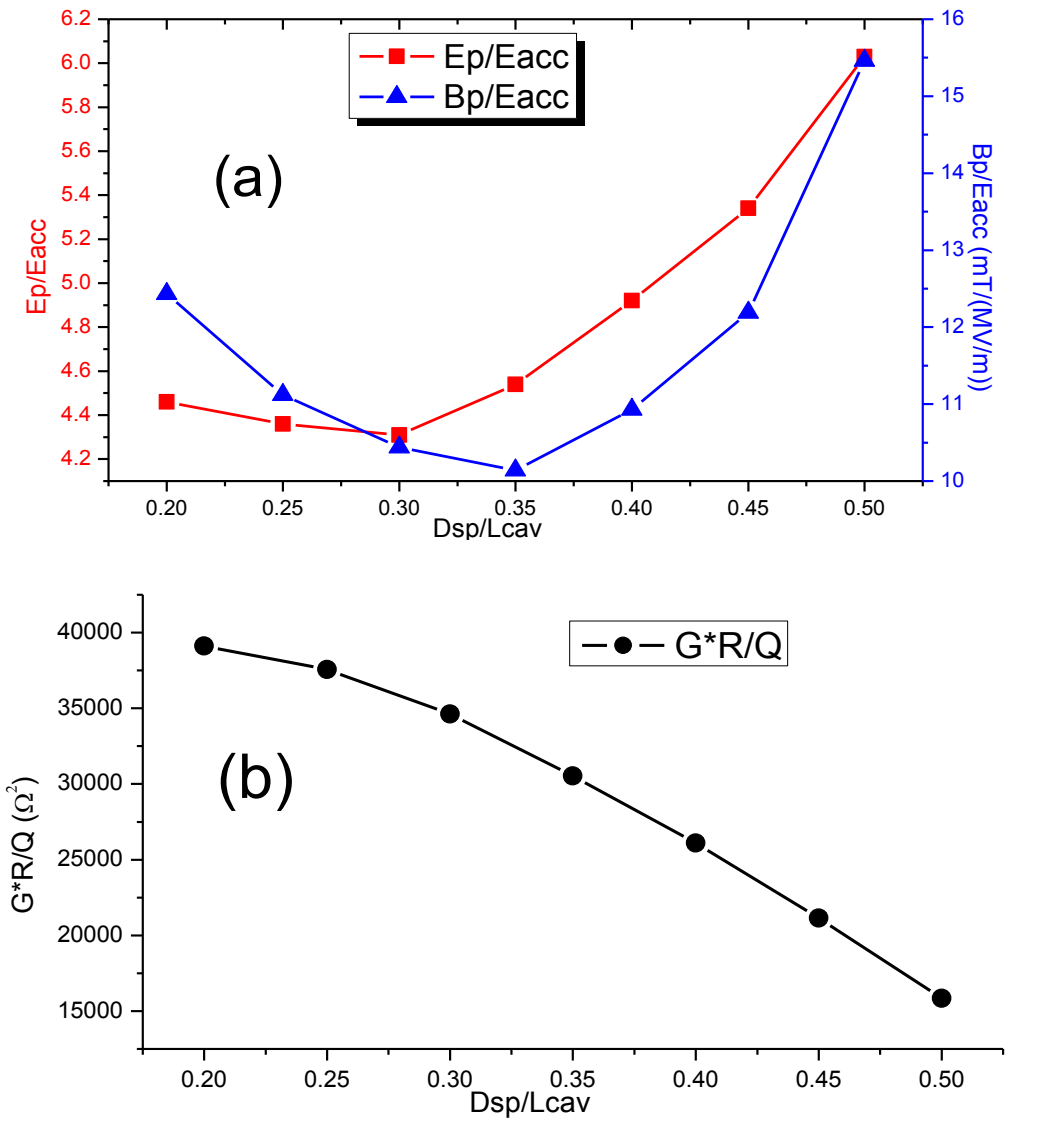

Fig.2. Dependence of (a)*Ep/Eacc* & *Bp/Eacc* and (b)$G*R/Q_0$ on *Dsp/Lcav*.

In figure 2, the *Bp/Eacc* and *Ep/Eacc* first decrease and then increase as the diameter of the spoke base increases. On the other hand, the $G*R/Q_0$ decreases as *Dsp* increases. Finally, we choose *Dsp/Lcav* as 0.36.

Then we fix the ratio of *Dsp/Lcav* and change the cavity length. Figure 3 shows the normalized field and shunt impedance change for a varying *Lcav*.

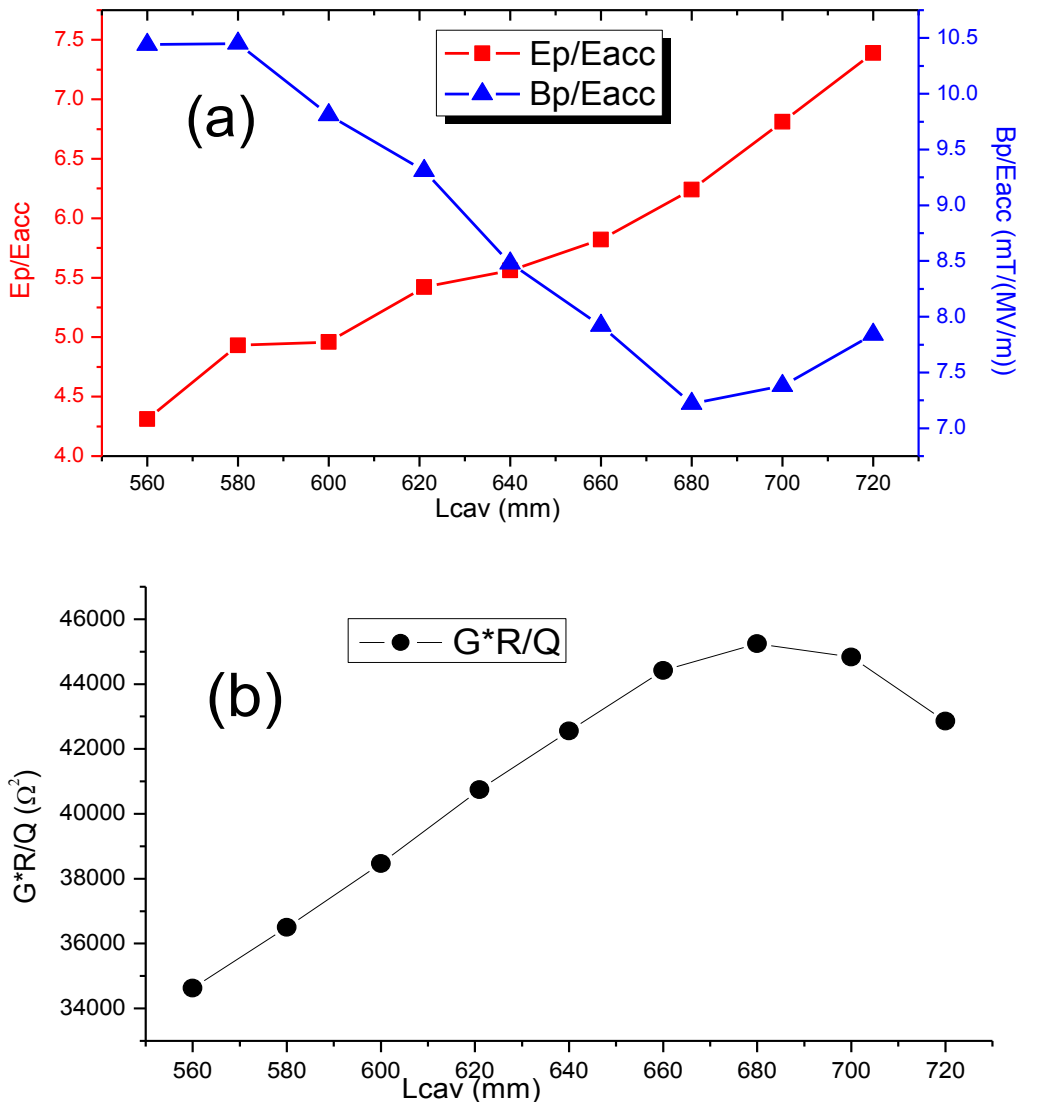

Fig.3. Dependence of (a)*Ep/Eacc* & *Bp/Eacc* and (b)$G*R/Q_0$ on *Lcav*.

In figure 3, it is clear that as the cavity length increases, the normalized magnetic field decreases and the shunt impedance increases whereas the normalized electric field increases. On the other hand, the iris-to-iris length is nearly a constant, so the length of cavity cannot be too large for mechanical stability and manufacturability.

We found that the surface magnetic field in figure 4(a) focused along the beam pipe with the cylindrical spoke base.

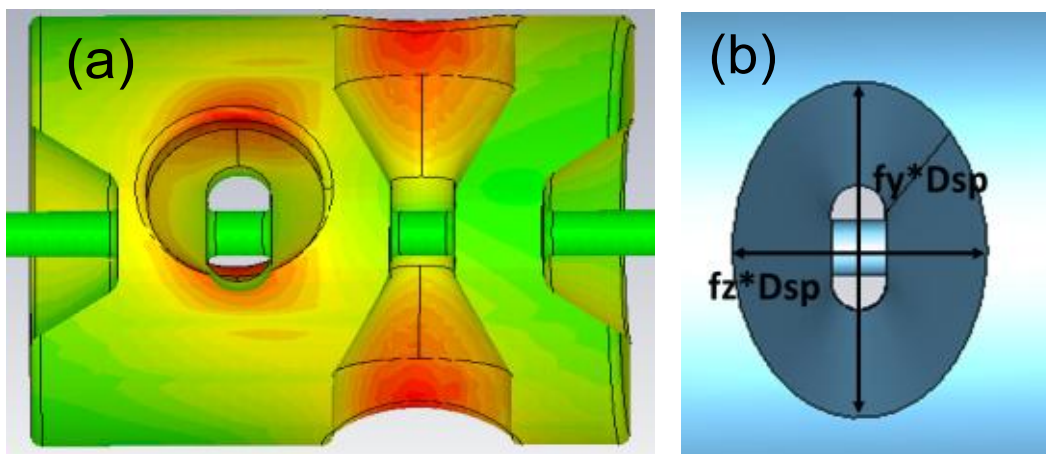

Fig.4. (a) view of the surface magnetic field with the cylindrical spoke base and (b) the elliptic column spoke base model and main geometric parameters, in which *Dsp* is fixed.

To uniform the surface magnetic field on the spoke base, we use the elliptic column (Figure 4(b)) instead of cylinder. We define a spoke base as being longitudinal if the longest dimension is parallel to the beam line and transverse otherwise, following Delayen [2]. Additionally, for convenience, the spoke base height *H* is zero, when varying the length and width of the spoke base.

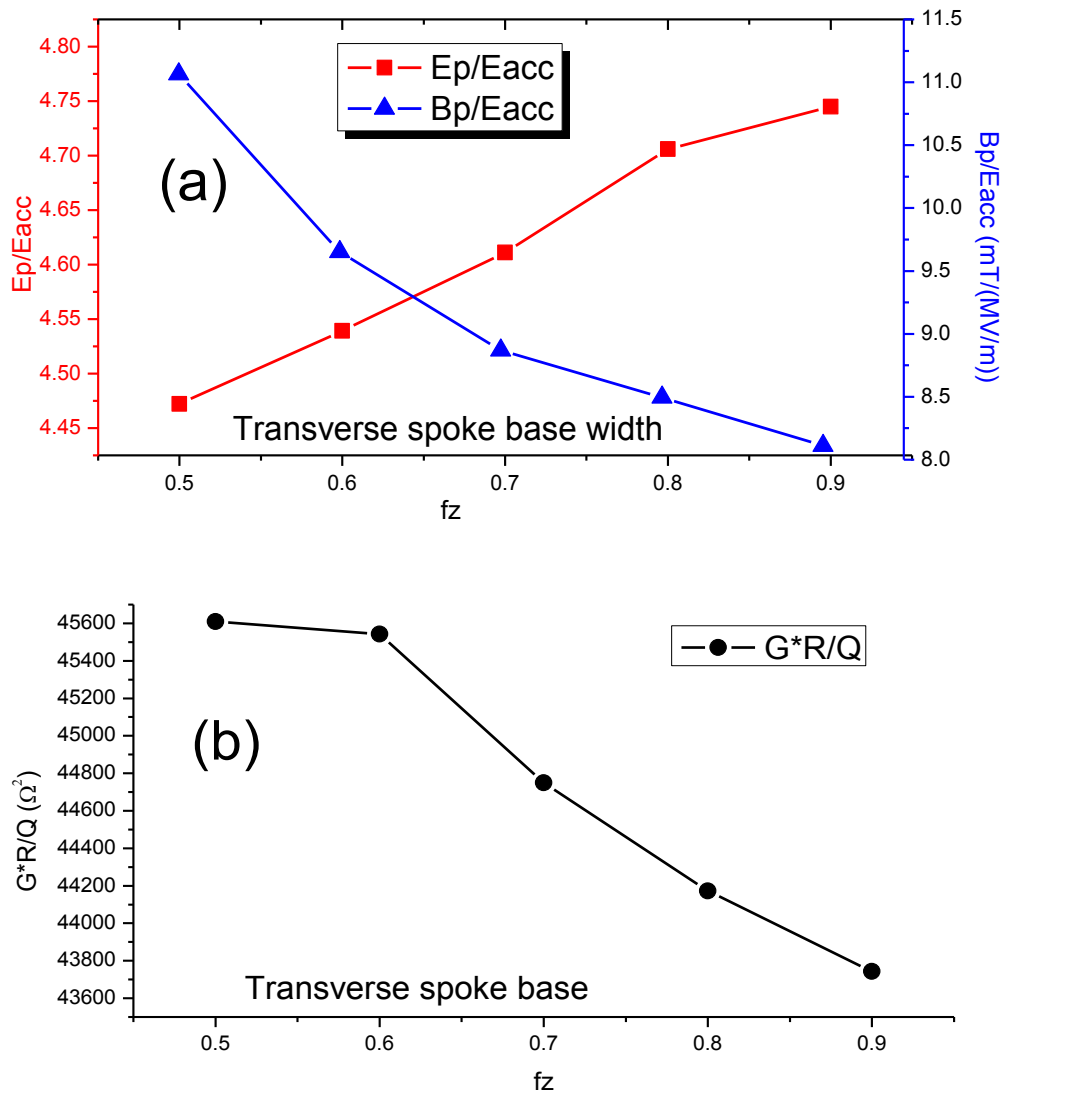

Fig.5. Dependence of (a)*Ep/Eacc* & *Bp/Eacc* and (b)*G*R/Q*$_0$ on *fz*Dsp*.

Figure 5 shows that increasing the transverse spoke base width (*fz*Dsp*) has adverse effects on both the normalized electric field and the shunt impedance.

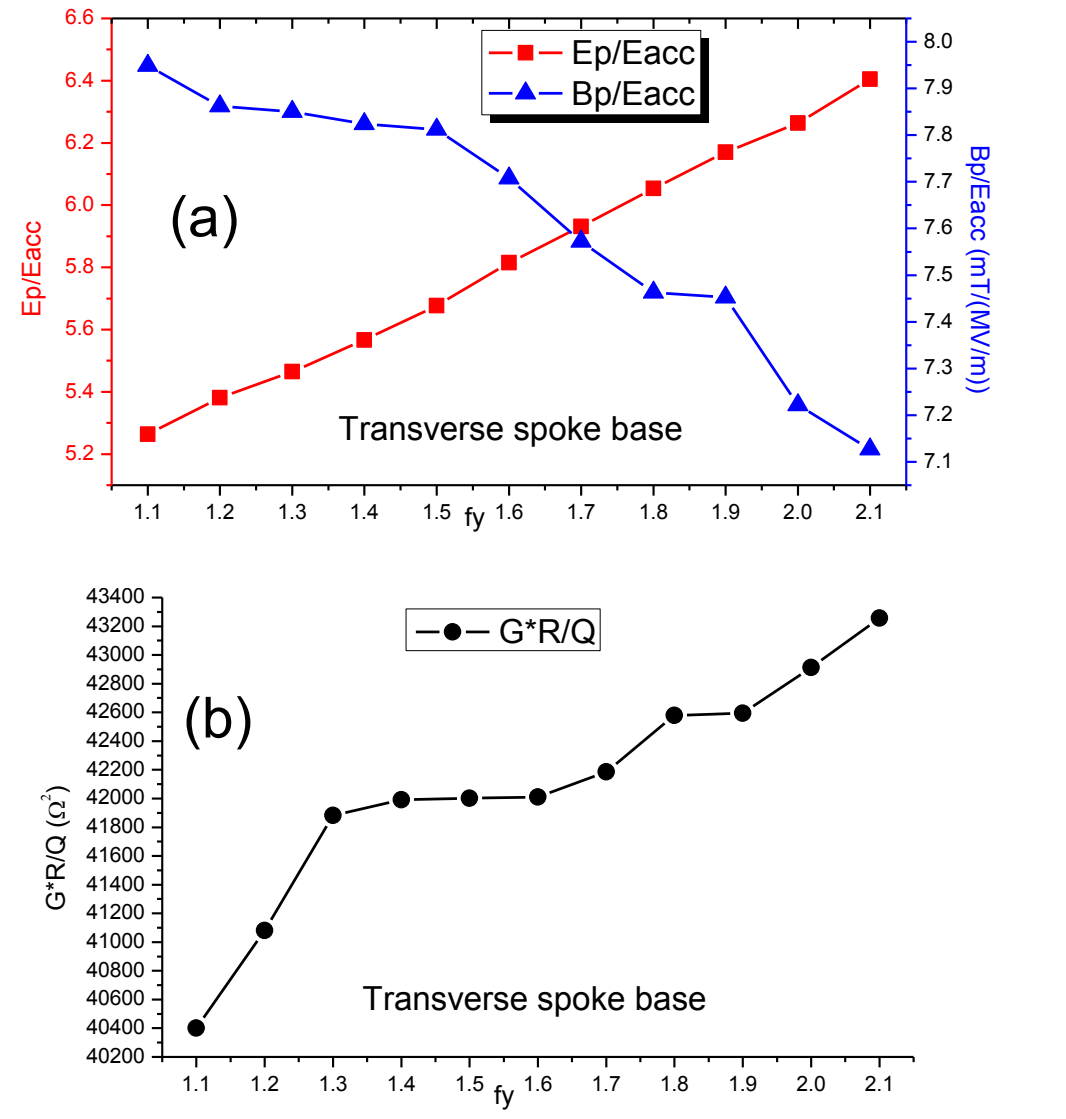

Fig.6. Dependence of (a)*Ep/Eacc* & *Bp/Eacc* and (b)*G*R/Q*$_0$ on *fy*Dsp*.

In figure 6, the transverse spoke base length (*fy*Dsp*) is varied. As it increases, the normalized magnetic field and the shunt impedance increase and the normalized electric field decreases.

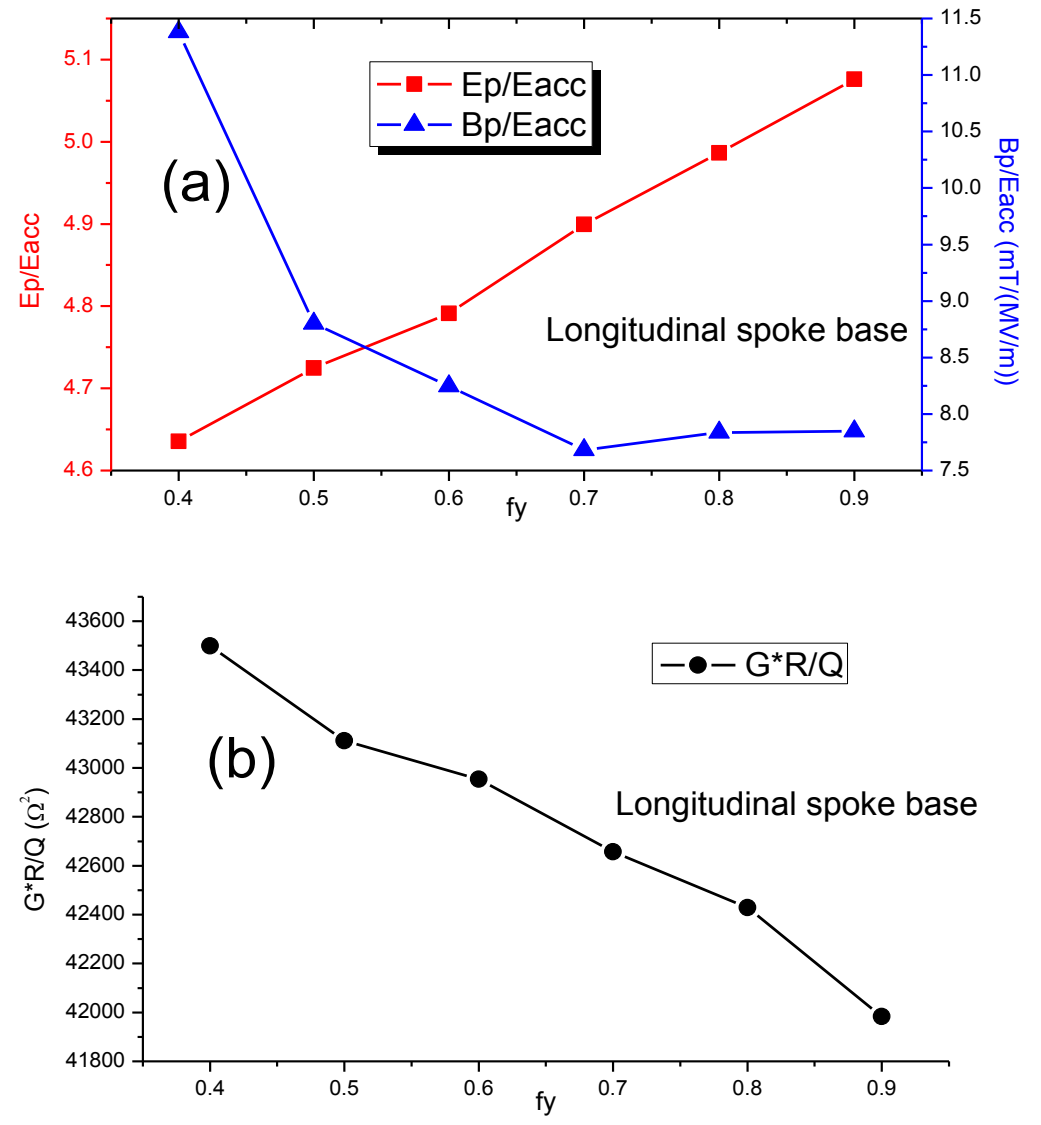

Fig.7. Dependence of (a)*Ep/Eacc* & *Bp/Eacc* and (b)*G*R/Q*$_0$ on *fy*Dsp*.

Figure 7 illustrates that increasing the longitudinal spoke base width (*fy*Dsp*) leads to a lower *Bp/Eacc* and a higher *Ep/Eacc* and *G*R/Q*.

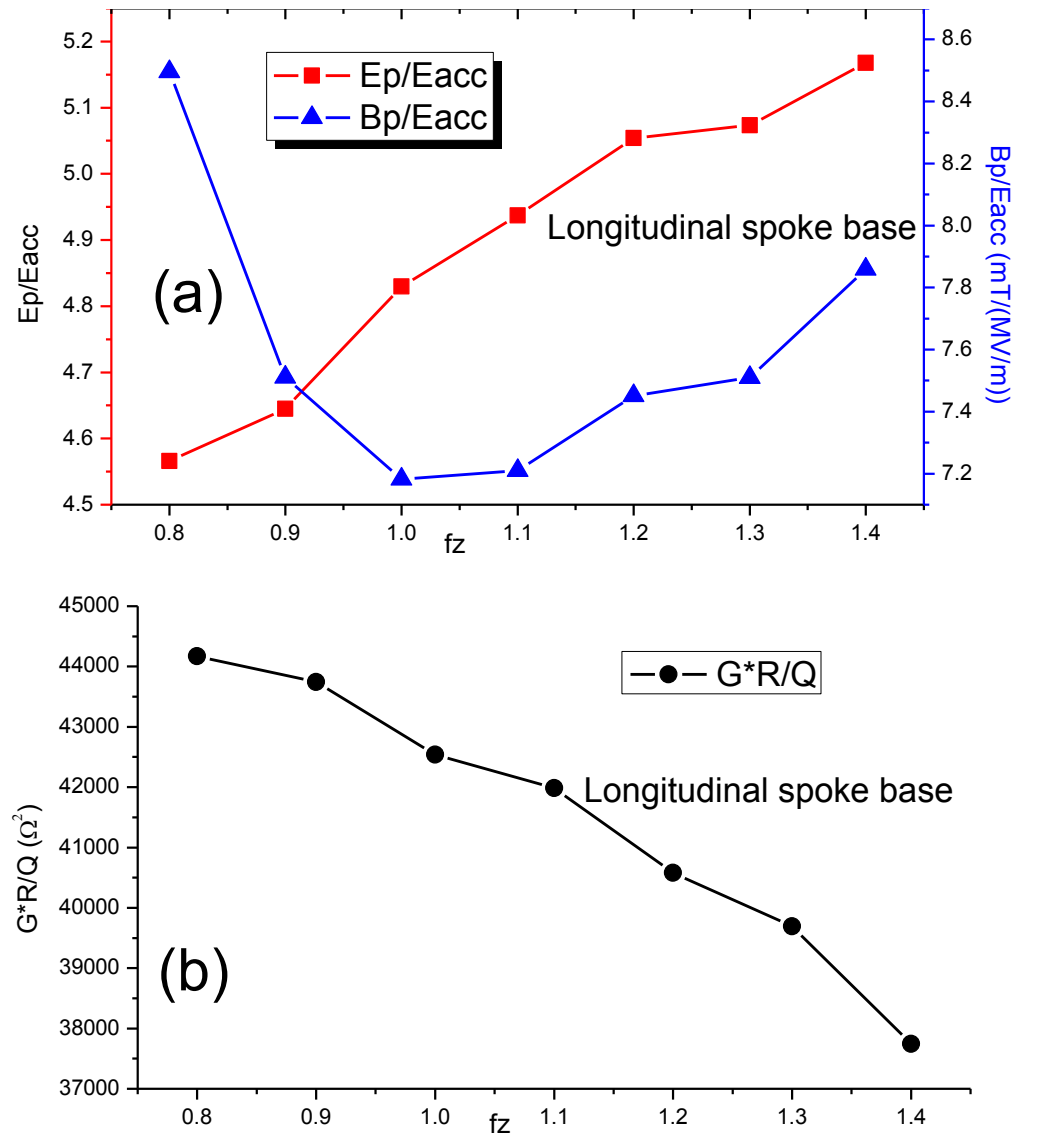

Fig.8. Dependence of (a)*Ep/Eacc* & *Bp/Eacc* and (b)*G*R/Q*$_0$ on *fz*Dsp*.

Figure 8 shows a parameter sweep of the longitudinal spoke base length (*fz*Dsp*) to show how the normalized fields and the shunt impedance are effected.

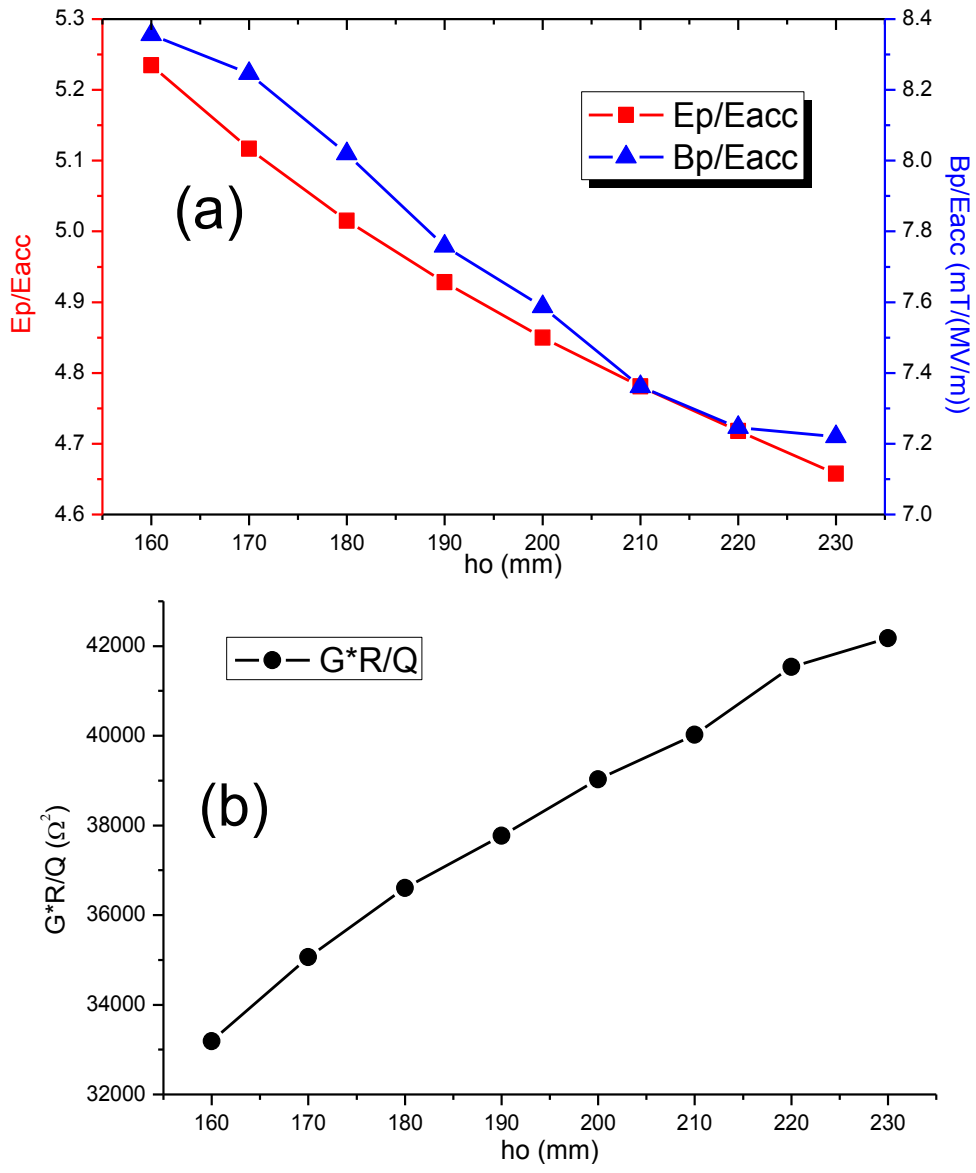

Fig.9. Dependence of (a)*Ep/Eacc* & *Bp/Eacc* and (b)$G*R/Q_0$ on ho.

The parameter ho, as can be seen in figure 1, is related to spoke base height *H*, which has an effect on both the normalized fields and shunt impedance. Figure 9 illustrates that the smaller spoke base height, the better RF properties.

Figure 7 and 8 show that a transverse spoke base has a lower normalized magnetic field and higher shunt impedance than a longitudinal spoke base. However, for a cylindrically shaped outer conductor, without spoke base height will become very difficult to fabricate and clean. Therefore, for easy machining, we use a longitudinal spoke base.

**Spoke aperture:** The electric field of the accelerating mode of a spoke cavity are concentrated at the spoke aperture region. To optimize the normalized field, we need to change the thickness (*t*), width (*w*) and height (*h*) of the racetrack shape spoke aperture.

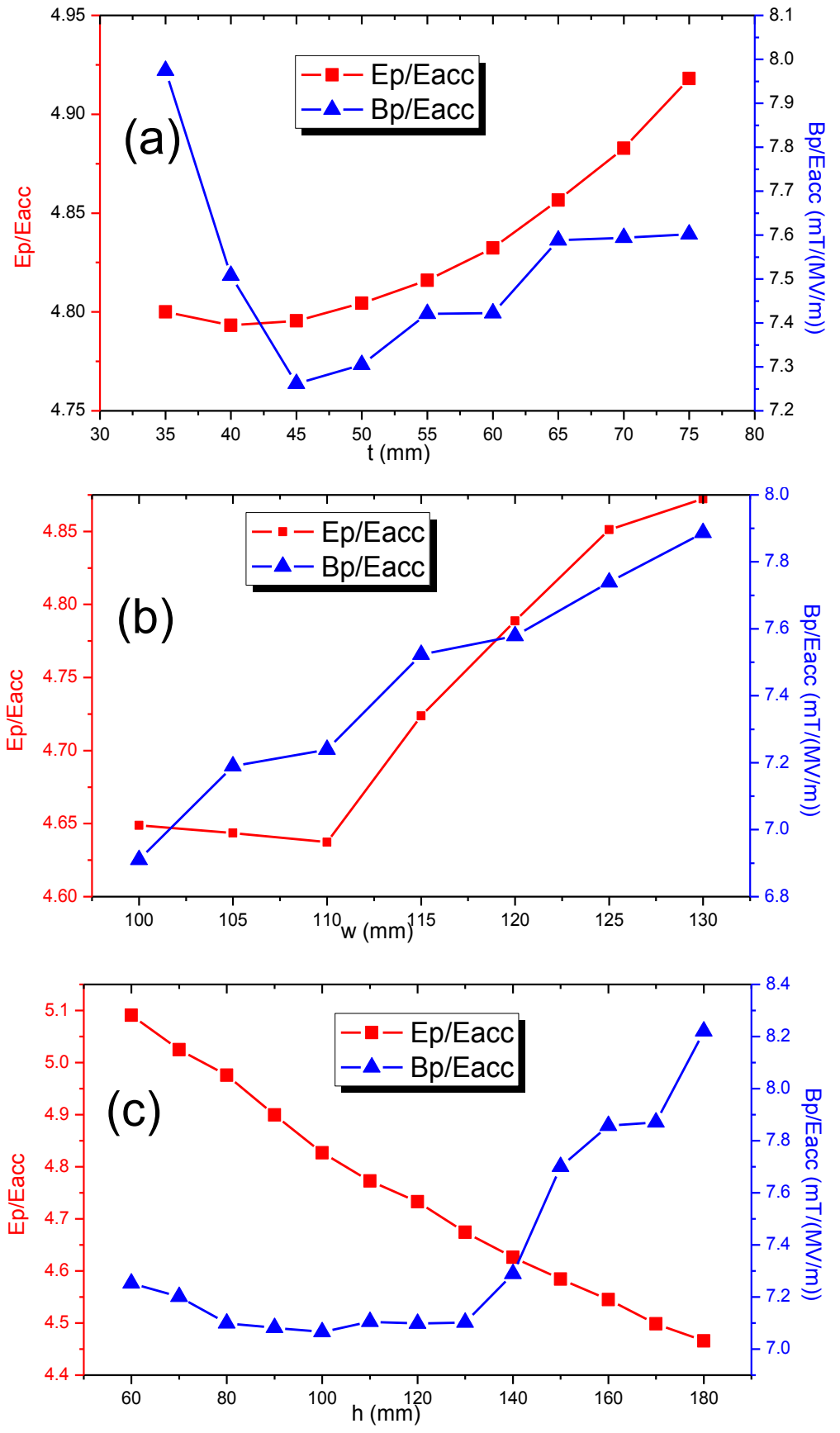

Fig.10. Dependence of *Ep/Eacc* & *Bp/Eacc* on (a)*t*, (b)*w* and (c)*h*.

In figure 10(a), a value of around 45mm sufficiently minimizes both the *Ep/Eacc* and the *Bp/Eacc*. It is also clear that in figure 10(b) as the width of the spoke aperture increases, the normalized field increases. Figure 10(c) shows that as the spoke aperture height increases, the normalized magnetic field decreases and the normalized electric field increases. Finally, we choose h as 138mm.

## 2.2 Coupler and clean ports

Four clean ports are added on the end wall to permit good cleaning access to the cavity. The position of the coupler and pick-up ports between the spoke base length axis and the end wall (figure 11) looks attractive because of the magnetic field in this region equals zero.

## 2.3 RF parameter

Table 1 summarizes all the RF parameters of the cavity design; the effective length is defined as $L_{eff} = 3\beta_0\lambda/2$. Figure 11 shows the electric and magnetic fields distribution in the double spoke RF volume with all the ports.

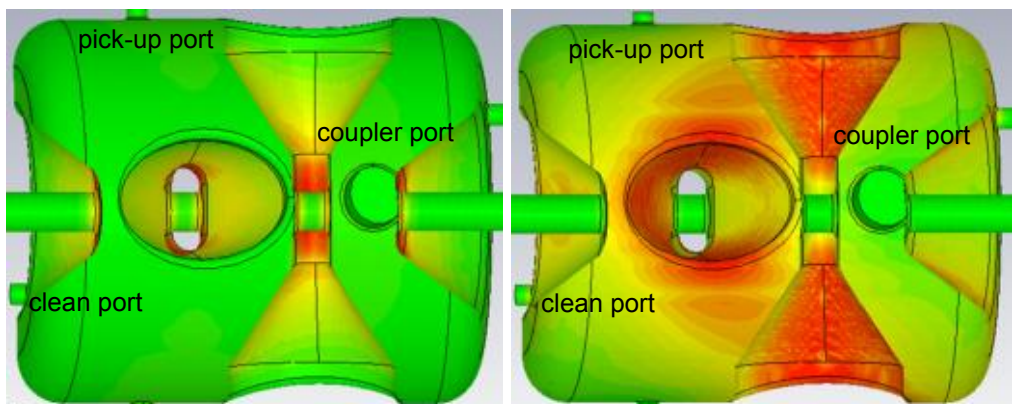

Fig.11. electric (left) and magnetic (right) field

Table 1. RF properties of the double spoke cavity, 325MHz, $\beta_0$=0.37.

| Parameter | Length | Diameter | Ep/Eacc | Bp/Eacc | $R/Q_0$ | G |
|---|---|---|---|---|---|---|
| Value | 620 [mm] | 493 [mm] | 4.39 | 7.15 [mT/MV/m] | 414.9 [Ω] | 108.6 [Ω] |

## 3 Mechanical studies

The cavity has been simulated under a one atmosphere external load, while both end beam pipes are fixed. According to the no stiffeners cavity result, we chose the stiffeners as figure 12 shown. All the stiffeners are 4 mm thick niobium. Table 2 gives the equivalent results of maximum Von Mises stress and displacement for both no stiffened and stiffened cavities. These stiffeners reduce the stresses at iris significantly and move the maximum stress point from the cavity body to the stiff ribs.

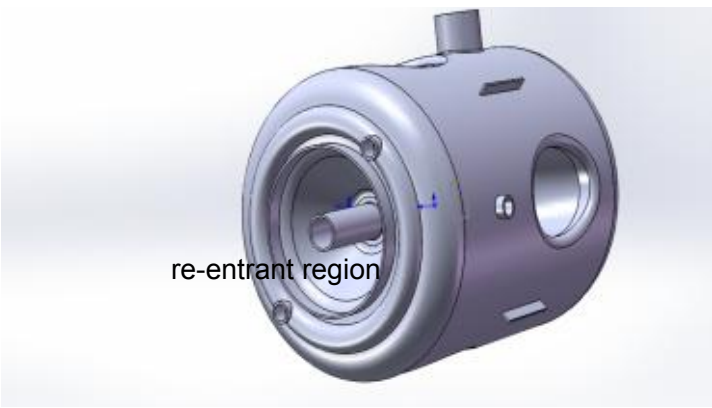

Fig.12. the cavity with stiffeners on the end wall

This flexibility in the re-entrant region also has benefit, which is used for reasonable deformations for cavity tuning. The tuning sensitivity and tuning force are shown in table 2 as well.

Table 2. Cavity mechanical properties

| | | |
|---|---|---|
| **1 atm pressure (without stiffeners)** | | |
| maximum displacement | 0.083 | mm |
| maximum stress Von Mises | 98.87 | MPa |
| rf frequency shift | -236 | kHz |
| **1 atm pressure (with stiffeners)** | | |
| maximum displacement | 0.045 | mm |
| maximum stress Von Mises | 42.15 | MPa |
| rf frequency shift | -24 | kHz |
| **tuning** | | |
| tuning sensitivity | 340 | kHz/mm |
| tuning force | 20 | kN/mm |

## 4 Conclusions

We presented the EM and mechanical study of a double spoke superconducting cavity intended for the China ADS. We use the elliptic column spoke base instead of cylindrical spoke base to decrease the normalized field. The optimization of the design provided values of Ep/Eacc and Bp/Eacc, which meets the requirement of China ADS. The fabrication will be under progress soon.

## References

1 M. Kelly. Status of Superconducting Spoke Cavity Development. Proc. SRF2007

2 C. S. Hopper, R. G. Olave, J. R. Delayen. Development of Spoke Cavities for High-Velocity Applications. Proc. IPAC2012

3 S. Nagaitsev, Project X – A New Multi-Megawatt Proton Source at Fermilab. Proc. PAC2011

4 S. Peggs. The European Spallation Source. Proc. IPAC2011

5 F. L. Krawczyk et al. Design of a Low-β, 2-Gap Spoke Resonator for the AAA Project. Proc. PAC2001

6 X. Wu et al. End-To-End Beam

Simulations for C-ADS Injector II. IPAC2013